\documentclass[twocolumn]{aastex631}
\usepackage{placeins}

\usepackage{amsmath,amssymb,amstext}
\usepackage{xcolor}
\usepackage{gensymb}
\usepackage{multirow}
\usepackage{rotating}
\usepackage{tabularx}
\usepackage{comment}
\usepackage[version=4]{mhchem} 
\usepackage{enumitem}
\setlist[itemize]{noitemsep, topsep=0pt, parsep=0pt, partopsep=0pt}

\defcitealias{LustigYaeger2023}{Lustig-Yaeger \& Fu et al. 2023}
\defcitealias{Moran2023}{Moran \& Stevenson et al. 2023}
\defcitealias{May2023}{May \& MacDonald et al. 2023}

\begin{document}

\title{JWST COMPASS: A NIRSpec G395H Transmission Spectrum of the Super-Earth GJ~357~b}

\author[0000-0002-4489-3168]{Jea Adams Redai} 
\altaffiliation{National Science Foundation Graduate Research Fellow}
\affiliation{Center for Astrophysics ${\ rm\mid}$ Harvard {\rm \&} Smithsonian, 60 Garden St, Cambridge, MA 02138, USA}

\author[0000-0002-0413-3308]{Nicholas Wogan}
\affiliation{NASA Ames Research Center, Moffett Field, CA 94035, USA}

\author[0000-0003-0354-0187]{Nicole L. Wallack}
\affiliation{Earth and Planets Laboratory, Carnegie Institution for Science, 5241 Broad Branch Road, NW, Washington, DC 20015, USA}

\author[0000-0003-4157-832X]{Munazza K. Alam}
\affiliation{Space Telescope Science Institute, 3700 San Martin Drive, Baltimore, MD 21218, USA}

\author[0000-0002-8949-5956]{Artyom Aguichine}
\affiliation{Department of Astronomy and Astrophysics, University of California, Santa Cruz, CA, USA}

\author[0000-0003-2862-6278]{Angie Wolfgang}
\affiliation{Eureka Scientific, 2452 Delmer Street Suite 100, Oakland, CA 94602-3017, USA}

\author[0000-0003-4328-3867]{Hannah R. Wakeford}
\affiliation{School of Physics, University of Bristol, HH Wills Physics Laboratory, Tyndall Avenue, Bristol BS8 1TL, UK}

\author[0009-0008-2801-5040]{Johanna Teske}
\affiliation{Earth and Planets Laboratory, Carnegie Institution for Science, 5241 Broad Branch Road, NW, Washington, DC 20015, USA}

\author[0000-0003-3623-7280]{Nicholas Scarsdale}
\affiliation{Department of Astronomy and Astrophysics, University of California, Santa Cruz, CA 95064, USA}

\author[0000-0002-6721-3284]{Sarah E. Moran}
\altaffiliation{NHFP Sagan Fellow} \affiliation{NASA Goddard Space Flight Center, 8800 Greenbelt Road, Greenbelt, MD 20771, USA}

\author[0000-0003-3204-8183]{Mercedes López–Morales}
\affiliation{Space Telescope Science Institute, 3700 San Martin Drive, Baltimore, MD 21218, USA}

\author[0000-0002-7500-7173]{Annabella Meech}
\affiliation{Center for Astrophysics ${\rm \mid}$ Harvard {\rm \&} Smithsonian, 60 Garden St, Cambridge, MA 02138, USA}

\author[0000-0002-8518-9601]{Peter Gao}
\affiliation{Earth and Planets Laboratory, Carnegie Institution for Science, 5241 Broad Branch Road, NW, Washington, DC 20015, USA}

\author[0009-0003-2576-9422]{Anna Gagnebin}
\affiliation{Department of Astronomy and Astrophysics, University of California, Santa Cruz, CA 95064, USA}

\author[0000-0003-1240-6844]{Natasha E. Batalha}
\affiliation{NASA Ames Research Center, Moffett Field, CA 94035, USA}

\author[0000-0002-7030-9519]{Natalie M. Batalha}
\affiliation{Department of Astronomy and Astrophysics, University of California, Santa Cruz, CA 95064, USA}

\author[0000-0001-8703-7751]{Lili Alderson}
\affiliation{Department of Astronomy, Cornell University, 122 Sciences Drive, Ithaca, NY 14853, USA}

\begin{abstract}

We present JWST NIRSpec/G395H transmission spectroscopy observations of GJ~357~b, a warm (\(T_{\mathrm{eq}} \approx 525\) K) super-Earth ($1.2\ \mathrm{R_{\oplus}} $, $1.84\ \mathrm{M_{\oplus}} $) orbiting a nearby M3-type star, with a median precision of 18 ppm and 27 ppm in NRS1 and NRS2, respectively. These precisions are obtained by binning the spectrum into 53 spectroscopic channels with a resolution of 60 pixels (around 0.02 $\mu$m) each. Our analysis of the transmission spectrum reveals no detectable atmospheric spectral features. By comparing the observed spectrum with 1D forward models, we rule out atmospheres with mean molecular weights (MMW) lower than 8 g/mol to $3 \sigma$, as well as atmospheres with metallicities less than 300x solar. The lack of a low MMW primary atmosphere is consistent with a primordial H$_2$ rich atmosphere having escaped, given the planet's $\gtrsim5$ Gyr age, relatively low surface gravity (log g = 3.09), and its likely history of substantial incident extreme ultraviolet radiation. 
We conclude that GJ~357~b most likely possesses either a high-MMW secondary atmosphere, perhaps rich in oxidized gases like CO\(_2\), or is a bare rock with no atmosphere. Upcoming scheduled JWST thermal emission observations could help distinguish between these scenarios by detecting signatures indicative of atmospheric heat redistribution or molecular absorption...

\end{abstract}



\section{Introduction}
\label{sec:intro}

The discovery of super-Earths—exoplanets with radii between 1 and 1.7~$R_{\oplus}$—opens a new frontier in astronomy (e.g.,  \citealt{Borucki2011}, \citealt{Batalha2013}), offering unique opportunities to study planets with no direct analogs in our solar system. These planets occupy the middle ground between Earth-like terrestrial planets and mini-Neptunes ($\geq \mathrm{1.8\ R_{\oplus}}$). They could host a wide range of atmospheres, including residual hydrogen-helium envelopes, thin high mean molecular weight (MMW) atmospheres, or they may have no atmosphere at all \citep{Rogers2010, Madhusudhan2019}. Understanding the atmospheric properties of super-earths offers critical insights into planetary formation and evolution in different stellar environments (e.g. \citealt{Zahnle2017}, \citealt{Heng2012}).

Despite super-earths being so numerous in the solar neighborhood (e.g., \citealt{Howard2012}; \citealt{Dressing2015}), detailed characterization of their atmospheres has remained elusive, hindered by observational constraints. Observations with the \textit{Hubble Space Telescope} and ground-based facilities have largely yielded featureless transmission spectra. These results rule out extended hydrogen/helium-dominated atmospheres, but the limited precision and spectral coverage of these facilities hinder the detection and characterization of higher MMW atmospheres \citep{DiamondLowe2018, Libby-Roberts2022}. 

The launch of JWST has ushered in a new era of exoplanet atmospheric characterization, offering unprecedented sensitivity and wavelength coverage (e.g., \citealt{Greene2016}). Its Near Infrared Spectrograph (NIRSpec) G395H grating, covering wavelengths from 2.87 to 5.14~$\mu$m at a resolving power of $R\sim2700$, is well-suited for detecting key molecular absorption features, particularly carbon species such as CO2 and CH4, in the atmospheres of small exoplanets (e.g., \citealt{Birkman2022}; \citealt{Jakobsen2022}). Importantly, JWST achieves instrumental noise floors as low as 5--20~ppm in transmission spectroscopy (e.g. \citealt{Rustamkulov2022}estimated pre-launch for NIRSpec PRISM, \citealt{LustigYaeger2023} measured for NIRSpec G395H), enabling the detection of atmospheric features in super-Earths that were previously inaccessible. To this end, early JWST observations have already begun to reveal that many super-Earths may have high-MMW atmospheres or none at all (e.g., \citetalias{LustigYaeger2023, May2023,Moran2023}; \citealt{Alam2024}; \citealt{Alderson2024}; \citealt{Kirk2024}; \citealt{Scarsdale2024}; \citealt{Alderson2025}). 

To better understand small exoplanet atmospheres, the JWST COMPASS (Compositions of Mini-Planets for Atmospheric Statistical Study) program (GO~2512; PIs: Batalha \& Teske) is conducting a systematic survey of eleven 1--3~$R_{\oplus}$ planets using NIRSpec/G395H transmission spectroscopy (\citealt{Batalha2022}). COMPASS targets were selected using a quantitative merit function (\citealt{batalha2023}) based on planet radius, insolation flux, stellar effective temperature, an observed mass measurement, and the expected precision achievable with NIRSpec/G395H. The program aims to explore the detectability and diversity of small exoplanet atmospheres, enabling population-level atmospheric constraints.

In this study, we focus on GJ~357~b (also known as TOI-562.01), a $1.22 \pm 0.08~R_{\oplus}$ and $1.84 \pm 0.08~M_{\oplus}$ rocky exoplanet orbiting a nearby M3V star. Discovered by the Transiting Exoplanet Survey Satellite (TESS) and confirmed through radial velocity measurements \citep{luque_planetary_2019, Jenkins}, it is one of the closest terrestrial exoplanets (at 9.44 pc) with both mass and radius measurements, making it an excellent candidate for atmospheric characterization. Furthermore, GJ~357  is notably quiescent, ranking among the least active M-dwarfs observed \citep{modirrousta-galian_gj_2020}, which implies a lower high-energy irradiation environment. This reduced activity may moderate atmospheric escape processes, setting the system apart from other M-dwarf hosts observed with JWST. With an equilibrium temperature of around 525K,  GJ~357~b -- being the innermost planet in its system -- is among the warmer super‑Earths orbiting M dwarfs, where enhanced stellar irradiation is expected to drive atmospheric escape that can significantly alter atmospheric composition (e.g., \cite{lugerbarnes2015}; \cite{Zahnle2017})

We present JWST NIRSpec/G395H transmission spectroscopy observations of GJ~357~b, aiming to detect and characterize its atmosphere. By comparing the measured transmission spectrum with atmospheric models we aim to place constraints on the planet atmosphere's composition. Doing so allows us to assess how this planet's atmosphere may have evolved under its stellar irradiation environment. This paper is organized as follows: In Section~\ref{sec:obs}, we describe the observations, and in Section~\ref{sec:data_reduction} the data reduction methods. In Section~\ref{sec:results}, we present the analysis of the transmission spectrum and discuss the atmospheric constraints derived from the data. In Section~\ref{sec:discussion}, we interpret our findings in the context of atmospheric escape and planetary evolution. Finally, in Section~\ref{sec:summary}, we summarize our conclusions and outline prospects for future observations.

%

\section{Observations and Data Reduction} 
\label{sec:obs}

\subsection{NIRSpec Observations}

We observed one transit of GJ~357~b with JWST/NIRSpec on UT 04 December 2023  05:24:15 - 10:15:18, using the 
G395H mode, which provides spectroscopy between 2.87--5.14\,$\mu$m across the NRS1 and NRS2 detectors (with a $\sim$0.1\,$\mu$m detector gap between 3.72--3.82\,$\mu$m). The observations were taken with the NIRSpec Bright Object Time Series (BOTS) mode using the SUB2048 subarray, the F290LP filter, the S1600A1 slit, and the NRSRAPID readout pattern. Our data consisted of 4401 integrations (divided into three segments) with 3 groups per integration, and was designed to be centered on the transit event with sufficient out-of-transit baseline.

\subsection{Data Reductions} 
\label{sec:data_reduction}

We reduced the data using two independent public pipelines -- Tiberius\ref{sec:tiberius} and Eureka! \ref{sec:Eureka} -- in line with previous NIRSpec/G395H transmission spectra analyses to verify the reliability of our results.

\begin{deluxetable}{cc}
\tablewidth{0pt}
\tablehead{\colhead{Stellar Parameters} & \colhead{Value}}
\startdata
M$_*$ ($M_\odot$) & 0.342 $\pm$ 0.011\tablenotemark{a} \\
R$_*$ ($R_\odot$) & 0.337 $\pm$ 0.015\tablenotemark{a} \\
P$_{\mathrm{rot}}$ (days) & 74.3 $\pm$ 1.7\tablenotemark{b} \\
log(g) & 4.94 $\pm$ 0.07\tablenotemark{a} \\
T$_{\mathrm{eff}}$ (K) & 3505 $\pm$ 51\tablenotemark{a} \\
$[\mathrm{Fe/H}]$ & -0.12 $\pm$ 0.16\tablenotemark{a} \\
K$_S$ (mag) & 6.475 $\pm$ 0.017 \tablenotemark{c}    \\
Age (Gyr) & $\ge$5\tablenotemark{d}    \\
\cutinhead{Planetary Parameters}
Period (days) & 3.93072 $\pm$ 0.00008\tablenotemark{e} \\
Mass ($M_\oplus$) & 1.84 $\pm$ 0.31\tablenotemark{e} \\
Radius ($R_\oplus$) & 1.217 $\pm$ 0.084\tablenotemark{e} \\
Semi-Major Axis (AU) & 0.035 $\pm$ 0.002\tablenotemark{e} \\
$\rho$ (g$\mathrm{cm^{-3}}$) & 5.6 $\pm$ 1.7\tablenotemark{e} \\
Inclination (\textdegree) & 89.12 $\pm$ 0.37\tablenotemark{e} \\
T$_{\mathrm{eq}}$ (K) & 525 $\pm$ 11\tablenotemark{e} 
\tablecaption{System parameters for GJ~357~b.}
\enddata
\label{tab:sys_params}
\tablenotetext{a}{\cite{Schweitzer2019}}
\tablenotetext{b}{\cite{Suarez2015}}
\tablenotetext{c}{The Two Micron All Sky Survey (2MASS), \cite{2Mass}}
\tablenotetext{d}{Lower Limit from X-Ray Observations, \cite{modirrousta-galian_gj_2020}}
\tablenotetext{e}{\cite{luque_planetary_2019}, T$_{\mathrm{eq}}$ assumes zero-bond albedo}
\end{deluxetable}

\subsubsection{Tiberius}
\label{sec:tiberius}

\begin{figure*}
    \centering
    \includegraphics[width=\textwidth]{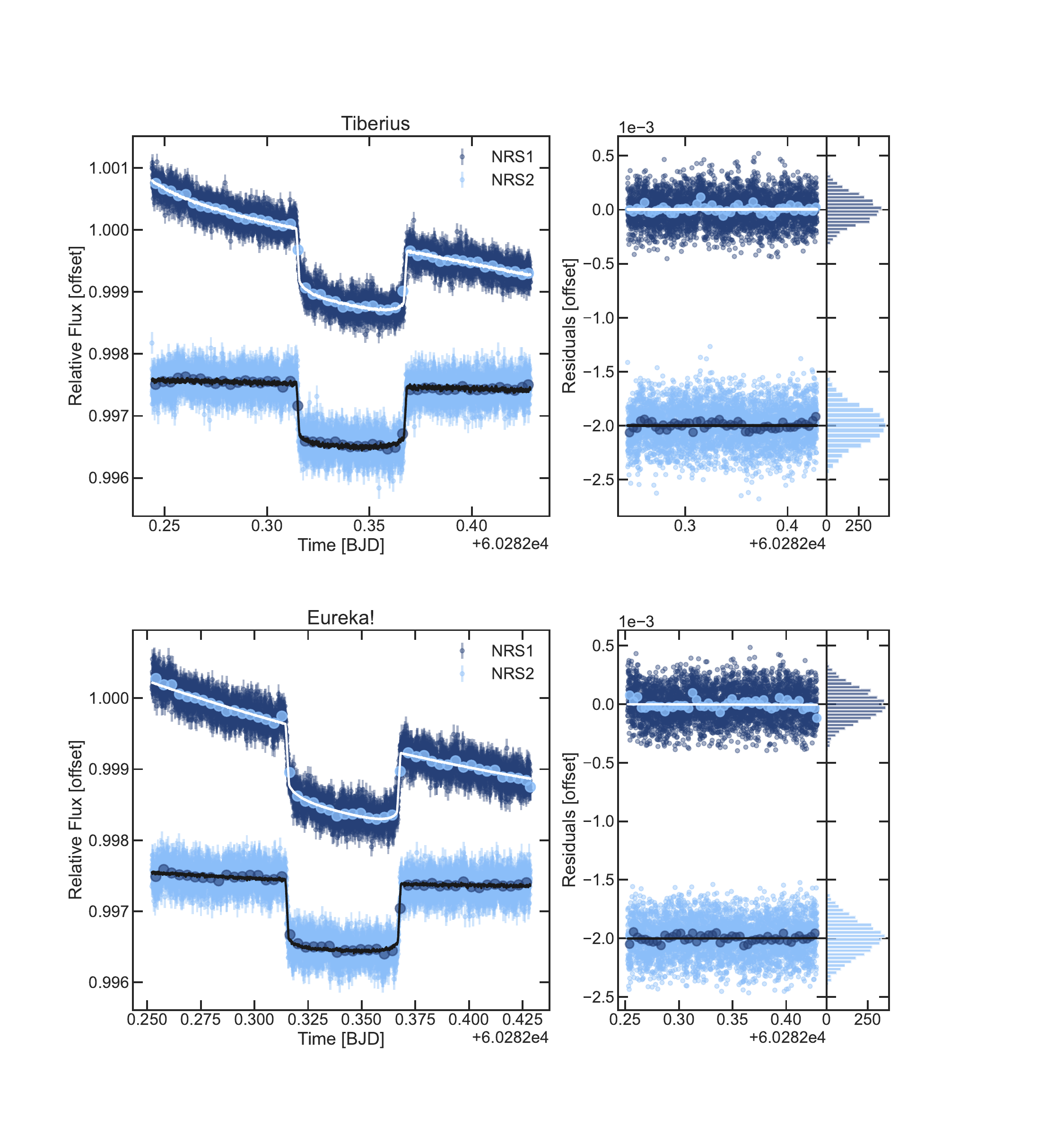} 
    \caption{Top: Tiberius white light curves for each detector and reduction. Our best-fit models, including systematics, are overplotted on the data in white (NRS1) and black (NRS2). Right column: the residuals from the models. Binned light curves and residuals are also shown. 
    Bottom: Same as top, for \texttt{Eureka!} Reduction}
    \label{fig:wlc_both}
\end{figure*}

We used the {\tt Tiberius} pipeline (\citealt[v1.0.0;]{Kirk2017,Kirk2021}), which has been used in several NIRSpec studies to date (e.g. \cite{alderson_early_2023, Moran2023, Kirk2024}). Our analysis of the GJ-527-b data starts by processing the \texttt{uncal.fits} files with Stage 1 of the \texttt{jwst} pipeline (version 1.8.2). We then apply a custom 1/$f$ correction (group-level background subtraction), which involves subtracting the median of each pixel column after masking pixels centered on the curved stellar traces. Next, we perform spectral extraction on the \texttt{gainscalestep.fits} files, and implement the \texttt{jump} step with the default rejection threshold of 10$\sigma$. We then run the \texttt{assign\_wcs} and \texttt{extract\_2d} steps of the \texttt{jwst} pipeline to retrieve the wavelength arrays.

After Stage 1, we created bad-pixel masks for each pixel by identifying hot pixels within the data, then combined them with 5$\sigma$ outlier pixels identified via running medians (calculated 
with a sliding 
of three pixels) operating along the pixel rows. Prior to identifying the spectral trace, we interpolated each column of the detectors onto a grid 10$\times$ finer than the initial spatial resolution. This step reduces the noise in the extracted data by improving the extraction of flux at the sub-pixel level, particularly where the edges of the photometric aperture bisect a pixel. We also interpolated over the bad pixels using only 1 pixel to each side of the bad pixel.

We traced the spectra by fitting Gaussians at each column and used a running median, calculated with a moving box with a width of five data points, to smooth the measured centers of the trace. We fitted these smoothed centers with a 4th-order polynomial, followed by standard aperture photometry. To remove residual background flux not captured by the group-level background subtraction, we fit a linear polynomial along each column, masking the stellar spectrum, defined by an aperture with a width of 10 pixels centered on the trace we found in the previous step, from this background fit. We also masked an additional four pixels on either side of the aperture so that the background was not fitting the wings of the stellar PSF, and we clipped any pixels within the background that deviated by more than $3\sigma$ from the mean for that particular column and frame. After removing the background in each column, the stellar spectra were then extracted by summing within the ten-pixel wide aperture. The uncertainties in the stellar spectra were pulled from the uncertainties from the \texttt{ERR} extension of the FITS files. This allowed us to produce white light and spectroscopic transit lights curves from the extracted stellar spectra sequence. The white light curves are integrated over 2.86--3.69\,$\mu$m for NRS1, and 3.82--5.06\,$\mu$m for NRS2. For the spectroscopic light curves, we integrated over 60 pixel bins (around $ 0.02$\,$\mu$m each) which covered the same wavelength range as the white light curves. This resulted in 21 spectroscopic light curves for NRS1 and 32 for NRS2, for a total of 53 bins.

We fit our broadband light curves with a transit+systematics model, where the systematics model, including an exponential ramp factor, is expressed as:

\begin{equation}
\mathrm{S} = c_{1}\ +\ (c_{2}\times \mathrm{T})\ +\ (c_{3}\times \mathrm{X})\ +\ (c_{4}\times \mathrm{Y})\ +\ (c_{5}\times \mathrm{F})  
\label{eq:1}
\end{equation}

\begin{equation}\label{eq:sys_model}
    \mathrm{Model} =  \mathrm{S} \times r1 \times \mathrm{exp}(r2 \times \mathrm{T})
\end{equation}

\noindent where T is time,  X and Y are the x- and y- pixel positional shifts
on the detector, F is the full width half maximum, and c1--c5 are the coefficients we fit for in the linear polynomial and r1 and r2 are free parameters in the exponential ramp. This fit was done using the Levenberg-Marquardt algorithm. These white light curves are fitted for $R_{p}/R_{\star}$, $i$, $T_{0}$, $a/R_{\star}$,  and the coefficients of the systematics model parameters ($x$ and $y$ pixel shifts, FWHM, and background) with the broadband light curves (NRS1: 2.86--3.69\,$\mu$m, NRS2: 3.82--5.06\,$\mu$m) generated from the \texttt{Tiberius} stellar spectra. We fixed $P$ and the quadratic limb darkening coefficients $u_1$ and $u_2$ to the values given in Table \ref{tab:sys_params}, with the latter computed using the 3D Stagger grid \citep{Magic2015} via {\tt ExoTiC-LD} \citep{Grant2024}. Eccentricity, $e$, was set to zero \citep{luque_planetary_2019} and longitude of periastron, $\omega$, was set to ninety. We used uniform priors for all the fitted parameters. Our analytic transit light-curve model was generated with {\tt batman} \citep{Kreidberg2015}. Each fit is done in two iterations: The first iteration is used to rescale the photometric uncertainties to achieve $\chi^2_\nu = 1$, while the second iteration is run with the rescaled uncertainties. For the spectroscopic light curves, binned into 53 channels, we held $a/R_{\star}$, $i$, and $T_{0}$ fixed to the best-fit values from the broadband light curve fits given in Table \ref{tab:wlc_fits}. The {\tt Tiberius} white light curves and residuals are shown in the top panel of Figure \ref{fig:wlc_both}, and the best-fit white light curve parameters are shown in Table \ref{tab:wlc_fits}. We then fit the spectroscopic light curves, again using the Levenberg-Marquadt algorithm in two iterations, with a linear polynomial + exponential ramp systematics model as in Equation \ref{eq:sys_model}. We used wide bounds for all fitted parameters.

\subsubsection{Eureka!}\label{sec:Eureka}

We produced a second, independent reduction of the data with the open-source \texttt{Eureka!} pipeline \citep{2022JOSS....7.4503B}, as we have done in previous COMPASS papers (\citealt{Alam2024, Alderson2024, Scarsdale2024, Wallack2024, Alderson2025}; Teske \& Batalha, 2025 Accepted for Publication). In brief, we use version 0.10 of \texttt{Eureka!} as a wrapper for the first two stages of version 1.11.4 of the \texttt{jwst} pipeline (context map \texttt{jwst\_1225.pmap)}. Other than using a jump detection threshold of 15$\sigma$, we use the default \texttt{jwst} Stage 1 and Stage 2 steps and utilize the inbuilt \texttt{Eureka!} group-level background subtraction to account for the 1/$f$ noise. For Stage 3, we select the optimal extraction aperture, background aperture, sigma threshold for the outlier rejection during the optimal extraction, and polynomial order for an additional background subtraction by generating versions of the white light curves for all combination of extraction aperture half-widths between 4 and 8 pixels in 1 pixel steps, background aperture half-widths between 8 and 11 pixels in 1 pixel steps, sigma thresholds of 10 and 60, and either an additional full frame or column-by-column background subtraction and selecting the version of the reduction that minimized the median absolute deviation providing the lowest scatter. We select an extraction aperture half-width of 4 pixels for NRS1 and 5 pixels for NRS2, a background aperture half-width of 9 pixels for NRS1 and 8 pixels for NRS2, an additional full-frame background subtraction for NRS1 and an additional column-by-column background subtraction for NRS2, and a 10 sigma threshold for the optimal extraction for both NRS1 and NRS2. The white light curve for \texttt{Eureka!} is shown in the bottom panel of Figure \ref{fig:wlc_both}

We then extract 60-pixel wide spectroscopic light curves from 2.863–3.714 $\mu$m for NRS1 and 3.812–5.082 $\mu$m for NRS2. We again opt to not use the inbuilt \texttt{Eureka!} fitter as we have done in previous COMPASS papers, for increased flexibility. For the white light curves and spectroscopic light curves, prior to fitting, we iteratively trim 3$\sigma$ outliers from a 50-point rolling median three times and trim the initial 500 data points from the start of the observation to account for any initial settling. For our light curve fits, we treat the two detectors separately. For the white light curves, we fit for the $i$, $a/R_{\star}$, $T_{0}$, and $R_{p}/R_{\star}$ using \texttt{batman} \citep{Kreidberg2014}. Simultaneously, we fit a systematic noise model of the form 
\begin{equation}
S= p_{1} + p_{2}\times T + p_{3}\times T^{2} + p_{4}\times X + p_{5}\times Y , 
\end{equation}
where $p_{N}$ is a free parameter, $T$ is the array of times, and $X$ and $Y$ are arrays of the positions of the trace on the detector. We elect to include a quadratic time term in our noise model to be consistent with the preference with the \texttt{Tiberius} reduction for additional terms in time beyond just a linear function. Additionally, we fit for per-point error inflation term that is added in quadrature to the measured errors. We fix the quadratic limb darkening coefficients to the theoretical values computed with Set One of the MPS-ATLAS models with {\tt ExoTiC-LD} \citep{Grant2024}, assuming the stellar parameters in Table~\ref{tab:sys_params}. We utilize the Markov Chain Monte Carlo sampler {\tt emcee} \citep{Foreman-Mackey2013} to fit for our combined astrophysical and instrumental noise model, using 3$\times$ the number of free parameters for the number of walkers. We initialize the walkers using a Gaussian for each parameter centered at the best fit values from a Levenberg-Marquardt minimization. We discard an initial 50,000 step burn-in and run an additional 50,000 step production run. We take the median and standard deviation of the chains as the best fit values and their associated uncertainties respectively. For the spectroscopic light curves, we utilize the same procedure with the exception of fixing the $i$, $a/R_{\star}$, $T_{0}$ values for each bin to the best-fit values from the white light curves of their respective detector.

\begin{table*}[]
\centering
\caption{White light curve best-fit values for the {\tt Tiberius} and {\tt Eureka!} reductions.}
\label{tab:wlc_fits}
\begin{tabular}{c|c|c|c|c|c}
\hline \hline
Reduction & Detector & $T_{0}$ (MJD) &  $a/R_{\star}$ & $i$ (degrees) & $R_{p}/R_{\star}$ \\ \hline \hline
{\tt Tiberius} & NRS1   & 60282.3413$\pm$3e${-5}$  & 23.42$\pm$0.54   & 89.56$\pm$0.75 & 0.031328$\pm$0.000127 \\
 & NRS2   & 60282.3413$\pm$3e${-5}$ &23.85$\pm$0.11  & 90.01$\pm$ 0.35   &  0.030460 $\pm$0.000098  \\ \hline
{\tt Eureka!} & NRS1   & 60282.3413$\pm$3e${-5}$  & 22.23$\pm$1.00   & 89.02$\pm$0.42 & 0.03181$\pm$0.00012  \\
 & NRS2   & 60282.3413$\pm$3e${-5}$   & 23.75$\pm$0.33   &  90.00$\pm$0.37 & 0.03054$\pm$0.000088  \\ \hline
\end{tabular}
\end{table*}


\section{Interpretation of the GJ 357 \MakeLowercase{b} Transmission Spectrum} 
\label{sec:results}

\begin{figure*}
    \centering
    \includegraphics[width=0.99\textwidth]{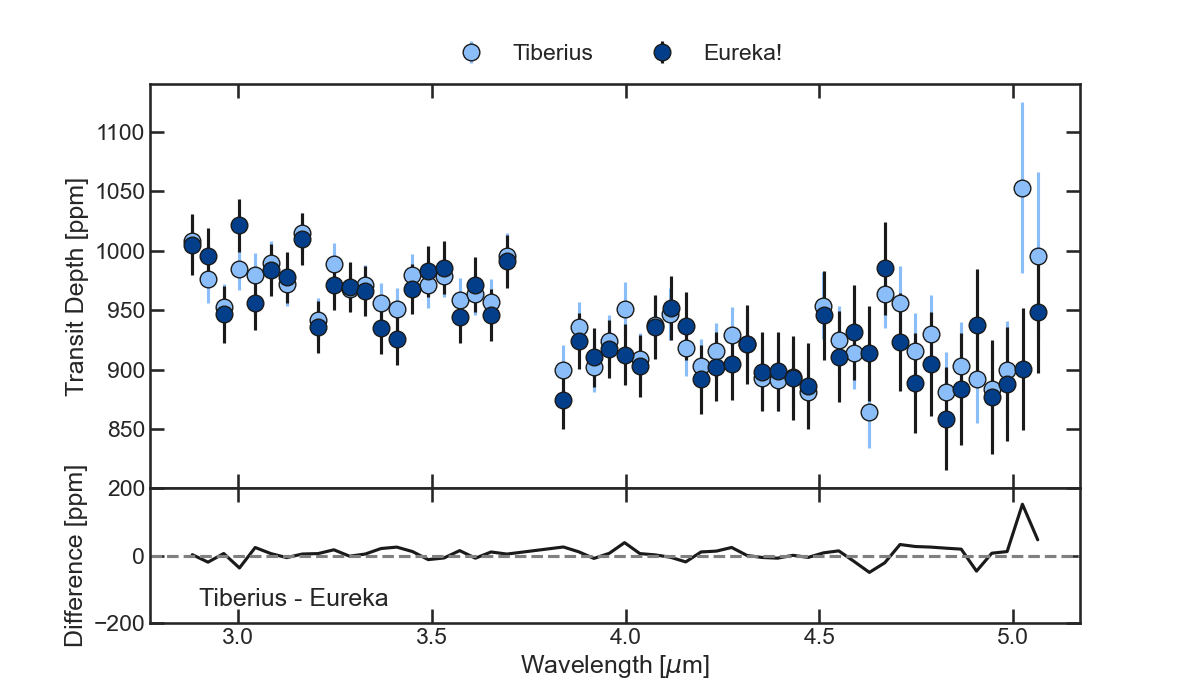} 
    \caption{Top: Comparison of the Tiberius (pale blue) and Eureka! (dark blue) transmission spectra of GJ~357~b in 60 pixel bins. Bottom: Difference between the two reductions.}
    \label{fig:ts}
\end{figure*}

Figure \ref{fig:ts} shows the 3–5~$\mu$m transmission spectrum of GJ~357~b, obtained using a 60-pixel binning scheme for our single visit. The spectra derived from the two independent reductions, {\tt Tiberius} and {\tt Eureka!}, show general consistency in shape. The pipelines were kept independent by design to ensure our conclusions were robust against data reduction choices.  All points in the spectrum agree within 3 sigma, and only one pair of points  fall outside each other's errorbars at 5.02$\mu$m. Such deviations at the edge of the detector can be due to differences in the aperture extraction radius and background region, the choice of aperture photometry in \texttt{Tiberius} versus optimal extraction in \texttt{Eureka!}, as well as the use of an exponential ramp in \texttt{Tiberius}.

\begin{figure*}
    \centering
    \includegraphics[width=0.99\textwidth]{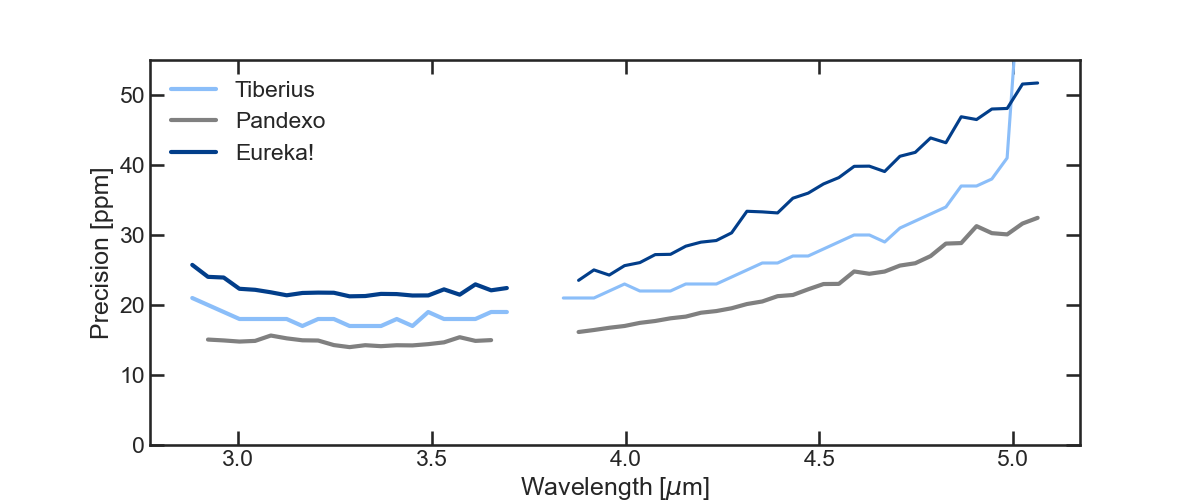} 
    \caption{Comparison of transit depth precisions achieved by each independent reduction (\texttt{Tiberius} in pale blue, \texttt{Eureka!} in dark blue) compared to the predicted values from \texttt{PandExo} simulations.}
    \label{fig:precision}
\end{figure*}

Below, to assess the atmospheric implications of the spectrum, we follow several previous works (e.g., \citetalias{Moran2023,May2023}; \citealt{Wallack2024,Alderson2024,Scarsdale2024}) by first searching for spectral features by fitting the data with non-physical models (Section \ref{sec:featuredet}). This analysis shows the spectrum is flat and featureless within the precision of our observations, which are 18 ppm and 27 ppm on average for detectors NRS1 and NRS2 respectively for Tiberius (Figure \ref{fig:precision}), and 22 ppm and 36 ppm for \texttt{Eureka!}'s NRS1 and NRS2 reductions. The Tiberius median precisions are 1.2$\times$ the expected Pandexo precisions across both detectors, and the \texttt{Eureka!} median precisions are 1.4$\times$ the Pandexo precisions in NRS1, and 1.6$\times$ Pandexo precisions in NRS2. While the Tiberius reduction is generally more precise than  \texttt{Eureka!}, it sees a spike in precision at the very edge of the detector. This is likely due to  limits in signal-to-noise ratio in this region, where \texttt{Eureka!}'s optimal extraction is able to define the aperture width to optimise SNR here. We also compare the data to physical atmospheric models to determine the range of compositions ruled out by the featureless spectrum.

\subsection{Feature detection with non-physical models} \label{sec:featuredet}

Upon visual inspection, the transmission spectrum (Figure \ref{fig:ts}) shows a clear offset between the NRS1 and NRS2 detectors but no obvious absorption features. To quantitatively confirm the ``flatness'' of the data, we fit  two non-physical models. The first model is two zero-slope lines, one for NRS1 wavelengths ($\lambda < 3.78$ $\mu$m) and the other for NRS2 (two total free parameters). This model captures any possible vertical offset between the NRS1 and NRS2 detectors. The second model has the same parameters as the first, but also includes a Gaussian with a variable amplitude, wavelength center, and standard deviation that approximates an agnostic spectral feature (five parameters). We fit both models to the reductions binned at 60-pixels using the \texttt{Ultranest} nested sampling algorithm \citep{Buchner2021} with wide uniform priors for all parameters. Table \ref{tab:featuredetect} shows the inferred detector offsets and Bayesian evidences, as well as the ``Feature detection confidence'', which is a Bayes factor given by the Gaussian model evidence divided by the flat model evidence. The Bayes factor is 0.49 for the \texttt{Tiberius} reduction and 0.68 for the \texttt{Eureka!} reduction, values that indicate preference for the flat line model. Furthermore, the simulations identify a $56^{+6}_{-6}$ and $57^{+8}_{-8}$ ppm detector offset for the \texttt{Tiberius} and \texttt{Eureka!} reductions, respectively. This type of offset between the NRS1 and NRS2 detectors is a known systematic feature in JWST/NIRSpec G395H observations that use a small number of groups per integration and has been noted in other results from the COMPASS survey \citep{Scarsdale2024, Alam2024, Wallack2024, Teske2025}. By treating the offset as a free parameter in our models, we ensure this instrumental effect does not bias our final atmospheric constraints. We also tested higher order polynomial terms in time, and linear limb darkening parameters in the \texttt{Tiberius} reduction, but this did not remove the offset.

Given that our non-physical model analysis prefers the flat model for both reductions (Table \ref{tab:featuredetect}) and the lack of clear features at wavelengths relevant to common atmospheric absorbers (e.g., the 4.3 $\mu$m CO$_2$ feature or 3.35 $\mu$m CH$_4$ feature), we conclude that the spectrum is featureless. With our flat spectrum, we are incapable of identifying the precise nature of GJ~357~b's atmosphere, but the data still rules out a wide range of atmospheric compositions, which we address in \S \ref{sec:atmo_models}.


\begin{table*}
  \caption{Feature detection with non-physical models}
  \label{tab:featuredetect}
  \begin{center}
  \begin{tabularx}{.95\linewidth}{
        >{\centering\arraybackslash}p{0.1\linewidth} 
        >{\centering\arraybackslash}p{0.1\linewidth} 
        >{\centering\arraybackslash}p{0.1\linewidth}
        >{\centering\arraybackslash}p{0.1\linewidth} 
        >{\centering\arraybackslash}p{0.1\linewidth}
        >{\centering\arraybackslash}p{0.15\linewidth}
        >{\centering\arraybackslash}p{0.18\linewidth}
    }
    \hline
    \hline
    \multirow{2}{*}{Reduction} & \multicolumn{2}{c}{Flat model} & \multicolumn{2}{c}{Gaussian model} & \multirow{2}{*}{\parbox{2cm}{\centering Feature Det. Factor$^c$}} & \multirow{2}{*}{\parbox{2cm}{\centering Feature Det. Significance}} \\
     & $\ln(Z)$$^a$ & Det. offset$^b$ & $\ln(Z)$$^a$ & Det. offset$^b$ & &  \\
    \hline
    {\tt Tiberius} & -30.9 & $56^{+6}_{-6}$ & -31.6 & $57^{+6}_{-7}$ & 0.49 & Prefers Flat model \\
    {\tt Eureka!} & -28.0 & $57^{+8}_{-8}$ & -28.4 & $57^{+8}_{-8}$ & 0.68 & Prefers Flat model \\
    \hline  
    \multicolumn{7}{p{0.95\linewidth}}{
      $^a$The natural log of the Bayesian evidence. 
      
      $^b$Offset between the NRS1 and NRS2 detectors in ppm and 68\% confidence interval.

      $^c$The evidence of the Gaussian model divided by the evidence of the Flat model.
    }
  \end{tabularx}
  \end{center}
\end{table*}

\subsection{Atmospheric constraints from physical models} \label{sec:atmo_models}

To determine the range of atmospheres compatible with our featureless spectrum, we compare it to a variety of physical atmospheric models. First, we consider compositions based on multiples of solar metallicity at chemical equilibrium. Metallicity is not the best parameterization for the atmospheric composition of a rocky world like GJ~357~b because any secondary atmosphere would be a product of outgassing, escape, and other geochemical processes. Atmospheres based on metallicity are instead best suited for gas-rich planets (e.g., Neptune- or Jupiter-like) which are expected to have hydrogen-rich primordial compositions. However, we regardless consider atmospheres based on metallicity for GJ~357~b because it facilitates comparison with previous work \citep[e.g.,][]{Scarsdale2024} and is a reasonable proxy for mean molecular weight. We compute chemical equilibrium with the \texttt{equilibrate} module of the \texttt{Photochem} software package \citep[v0.6.2,][]{Wogan2024photochem} using thermodynamics for H, C, O, N, S, and Cl species from \citet{Wogan2024}, assuming the \citet{Asplund2009} composition for the Sun (C/O = 0.55). The \texttt{equilibrate} solver is based on the similar \texttt{easyCHEM} package \citep{Lei2024}. Calculations account for a variable surface pressure or a cloud deck with an ``opaque pressure level'' following previous research \citep[e.g.,][]{Scarsdale2024}. Additionally, we use a simple analytical pressure-temperature profile \citep{Guillot2010}. Finally, we compute the transmission spectrum of modeled atmospheres using \texttt{PICASO} \citep{Batalha2019} with opacities archived on Zenodo \citep{Batalha2022}.

We compare simulated spectra to the data with two approaches. One method fits the modeled spectrum to the data, permitting a detector offset, then does a simple $\chi^2$ test to estimate the likelihood that the difference between the observations and the model are produced purely by chance. We performed this test for metallicities between 1$\times$ and $1000 \times$ solar (steps of 0.1 in $\log_{10}$ space), and ``opaque pressure levels'' between 1 and $10^{-5}$ bar (steps of 0.5 in $\log_{10}$ space). The second method takes a Baysian approach and fits the model to the data using \texttt{Ultranest} considering variable metallicity, opaque pressure level, vertical offsets for each detector, and additionally an error inflation term (five total parameters). Following \citet{Line2015}, we add the $\log_{10}$ error inflation term ($f$) with the observed error ($\sigma$) in quadrature: 

\begin{equation}
s_i^2 = \sigma^2 + (10^f)^2,
\end{equation}
resulting in the log-likelihood function,

\begin{equation}
\ln \mathcal{L} = -\frac{1}{2} \sum_{i=1}^{N} \left[ \frac{(y_i - F_i)^2}{s_i^2} + \ln(2\pi s_i^2) \right].
\end{equation}
Here, $y_i$ and $F_i$ are the observed and modeled transit depth, respectively, at wavelength index $i$. We adopt wide uniform or log-uniform priors for all five parameters (e.g., 1 to $1000 \times$ solar metallicity and 1 to $10^{-5}$ bar ``opaque pressure level'').

\begin{figure*}
  \centering
  \includegraphics[width=\textwidth]{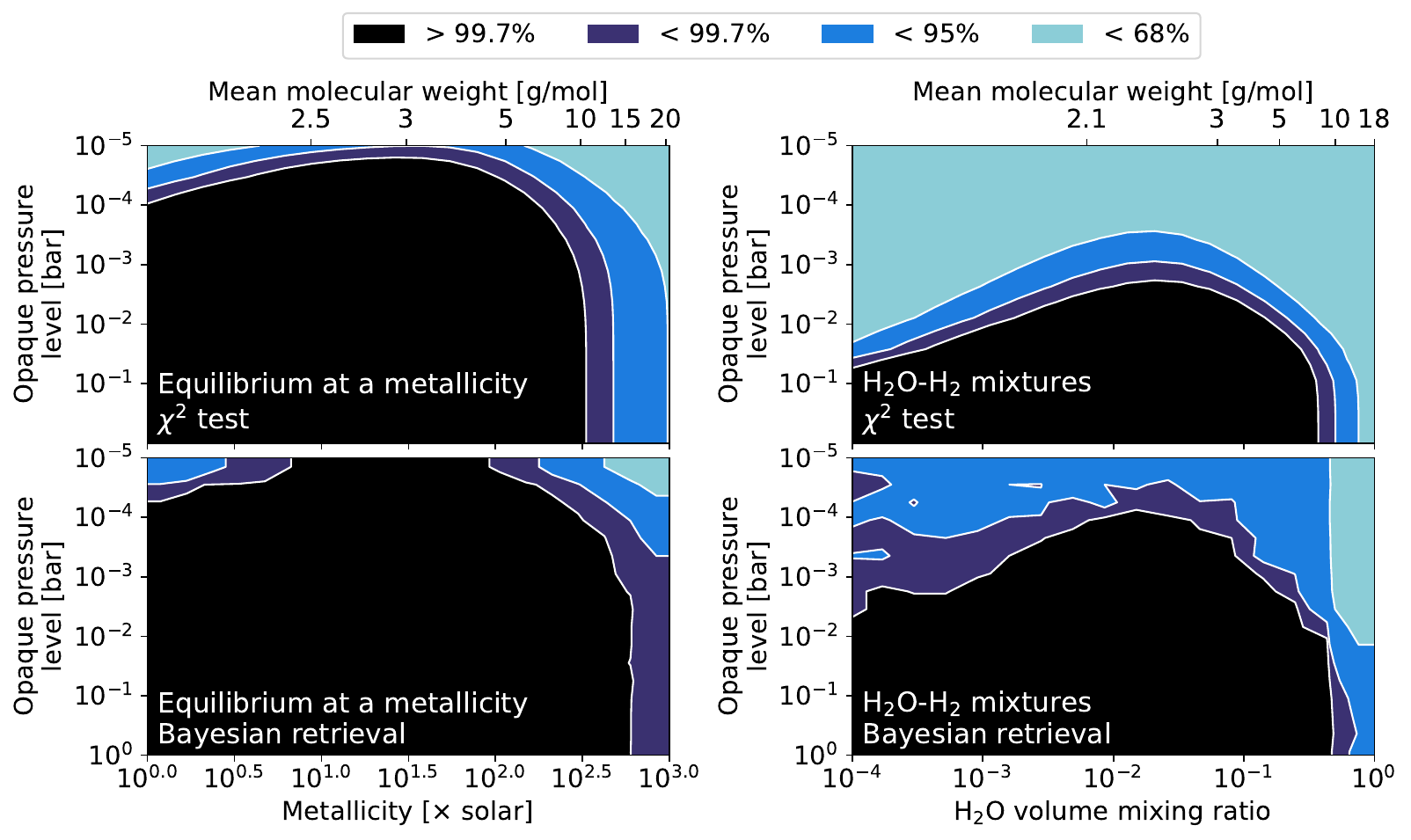}
  \caption{The atmospheric compositions ruled out by the \texttt{Tiberius} dataset binned at 60 pixels. In all sub-figures, the y-axis indicates the pressure at the top of an infinitely opaque layer (i.e., the surface or a cloud deck). Panels on the left consider atmospheres at chemical equilibrium at a given metallicity, while the panels on the right are simple H$_2$O-H$_2$ mixtures. Top panels show the statistical significance with which one can reject the null hypothesis that the difference between the observations and the model are produced purely by chance using a simple $\chi^2$ test, and the bottom panels provide the 1, 2, and 3-$\sigma$ contours of the posterior probability based on Bayesian retrievals. The black region is ruled out to $>99.7\%$ confidence (i.e., $>3\sigma$). To this high confidence, the JWST observations rule out metallicities $<10^{2.5} \times$ solar (i.e., MMW $\lesssim 10$) for opaque pressure levels $\gtrsim 10^{-4}$ bar regardless of the model-data comparison method (left). For H$_2$O-H$_2$ atmospheres (right) with opaque pressure levels $\gtrsim 10^{-1}$ bar, the data rule out MMW $\lesssim 8$.}
  \label{fig:ruleout}
\end{figure*}

The top left and bottom left panels of Figure \ref{fig:ruleout} show the range of metallicities and cloud top pressures excluded by the \texttt{Tiberius} reduction binned at 60 pixels for the $\chi^2$ test and Bayesian retrieval, respectively. The black regions are models ruled out to $>99.7\%$ confidence (i.e., $> 3\sigma$), which includes metallicities $\lesssim 300 - 500 \times$ solar, depending on the model-data comparison approach, for opaque pressure levels $\lesssim 10^{-4}$ bar. These threshold metallicities correspond to mean molecular weights between $10 - 15$ g/mol. The same analysis with the \texttt{Eureka!} reduction rules out the same parameter space: $\lesssim 300 - 500 \times$ solar metallicity for opaque pressure levels $\lesssim 10^{-4}$ bar. This result confirms that our conclusions are robust and not dependent on the specific reduction pipeline employed, despite the modest differences in the resulting precision.

We additionally used $\chi^2$ tests and a Bayesian retrieval to compare the \texttt{Tiberius} data to atmospheres composed of simple H$_2$O-H$_2$ mixtures. We choose H$_2$O-H$_2$ atmospheres (instead of e.g., CO$_2$-H$_2$ atmospheres) because this gas mixture has small amplitude and slowly varying spectral features in the NIRSpec G395H bandpass that should provide a conservative estimate for the MMW ruled out by the data. For the retrieval, we use the center-log-ratio transformation \citep{Benneke2012} to apply identical priors to H$_2$O and H$_2$, to prevent a bias towards either gas. The right two panels of Figure \ref{fig:ruleout} show the H$_2$O-H$_2$ mixtures ruled out by the \texttt{Tiberius} dataset using both model-data comparison approaches. For a $1$ bar opaque pressure level, the data exclude H$_2$O mixing ratios $\lesssim 40\%$ or MMW $\lesssim 8$ g/mol to $3\sigma$ for both the $\chi^2$ test and Bayesian retrieval. The retrieval finds that these H$_2$O concentrations are ruled out for opaque pressure levels $\gtrsim 10^{-3} - 10^{-2}$ bar, and the $\chi^2$ test requires the opaque pressure level $\gtrsim 10^{-2} - 10^{-1}$ bar to exclude the same range of H2O mixing ratios. These results are unchanged when we instead use the \texttt{Eureka!} dataset.

\begin{figure*}
  \centering
  \includegraphics[width=.99\textwidth]{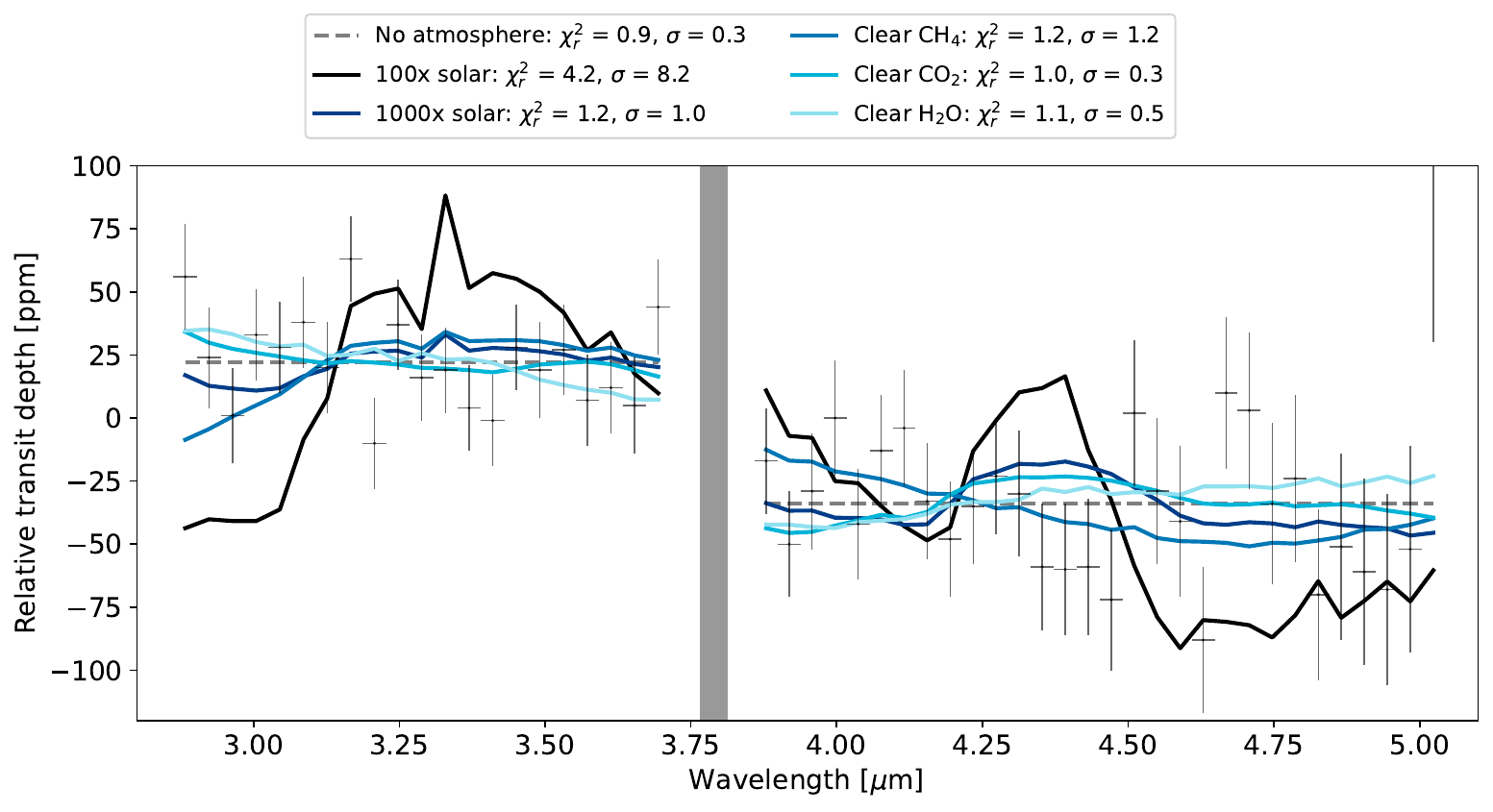}
  \caption{The \texttt{Tiberius} reduction compared to several atmospheric compositions for GJ~357~b. Modeled atmospheres are fit to the the NRS1 and NRS2 datasets independently, as indicated by the vertical grey bar. The $100 \times$ solar metallicity atmosphere is confidently ruled out by the data. All other compositions ($1000 \times$ solar metallicity, CH$_4$, CO$_2$, H$_2$O), are all permitted by the observations to within $1.2\sigma$. The data is also compatible with a bare rock with no atmosphere.}
  \label{fig:atm_examples}
\end{figure*}

Overall, if GJ~357~b has an atmosphere, our featureless spectrum suggests the atmosphere has a high MMW ($\gtrsim 8$ g/mol), consistent with other JWST transmission observations of warm rocky planets orbiting M stars \citep{Scarsdale2024,LustigYaeger2023,May2023,Moran2023}. A low MMW atmosphere remains possible only if the atmosphere has a high altitude cloud or haze layer (e.g., $< 10^{-2}$ bar; Figure \ref{fig:ruleout}). Figure \ref{fig:atm_examples} shows a variety of high MMW atmospheres compatible with our \texttt{Tiberius} spectrum to within $1.2\sigma$, including clear CH$_4$, CO$_2$, and H$_2$O compositions, and an atmosphere at chemical equilibrium with $1000\times$ solar metallicity. While any of these scenarios could explain the observations, an atmosphere is not required to fit the flat spectrum -- the data can easily be reconciled with a bare rock with no atmosphere at all ($0.3\sigma$).

\section{Discussion} 
\label{sec:discussion}

\subsection{GJ~357~b does not have a Low-MMW Primary Atmosphere} \label{sec:no_primary_atmopshere}

\begin{figure*}
  \centering
  \includegraphics[width=.99\textwidth]{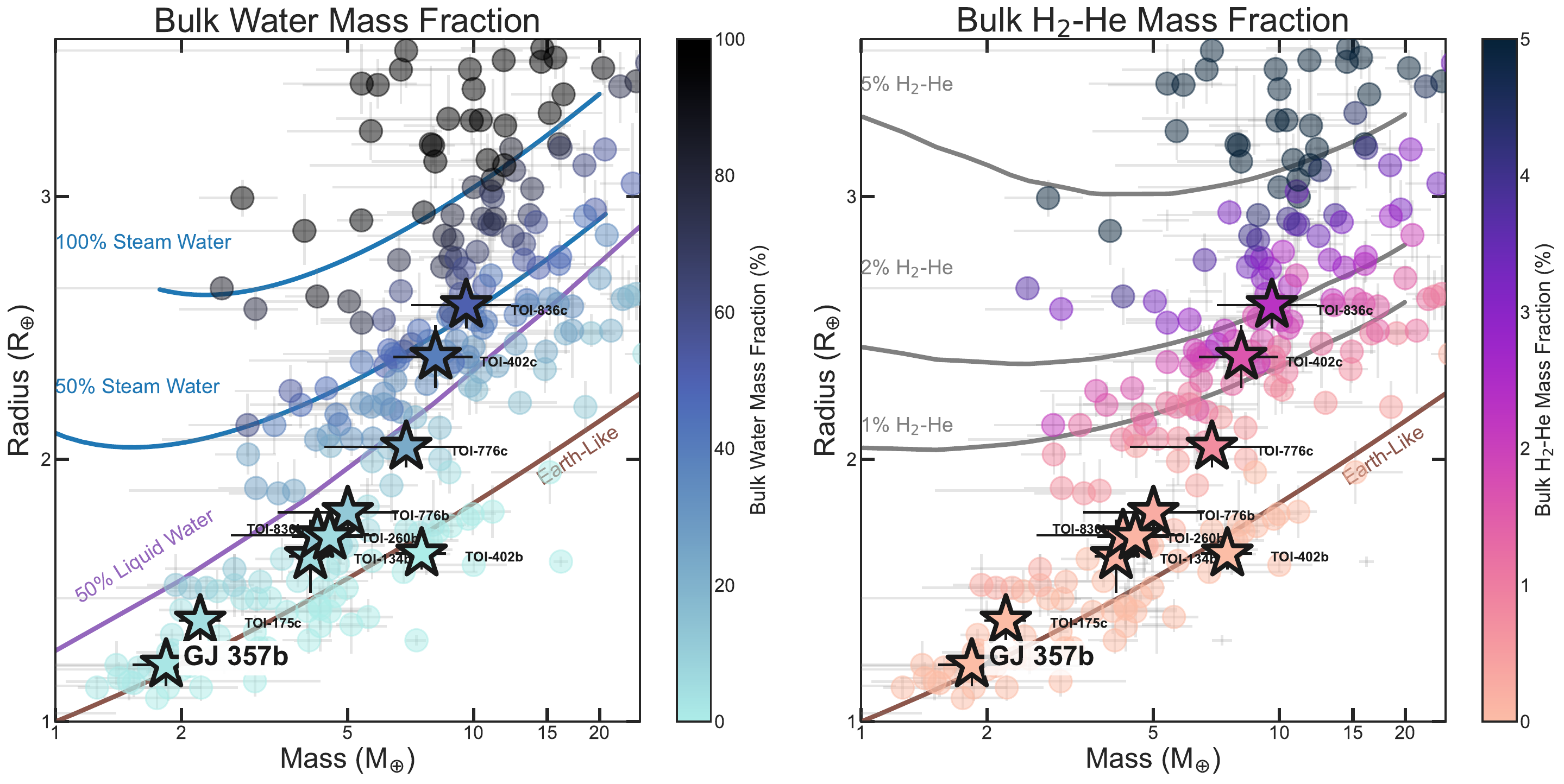}
  \caption{Mass-radius diagram for small planets ($0.2 \leq M_p \leq 30 M_\oplus$) derived from the NASA Exoplanet Archive, with planets shown as points and color-coded by composition. The left panel depicts the bulk water mass fraction \citep{Aguichine2021}, while the right panel shows the bulk H$_2$/He mass fraction \citep{Lopez2014}. Solid lines represent theoretical models of planetary composition at an equilibrium temperature of 500~K, including rocky Earth-like compositions from \citet{Zeng2016}. Bold, labeled stars mark other published COMPASS targets. GJ 357b is highlighted with a larger symbol and label.}
  \label{fig:mr_artem}
\end{figure*}


The absence of a low mean molecular weight atmosphere on GJ~357~b is consistent with its bulk density and the short lifetime of such an atmosphere due to escape processes. Figure \ref{fig:mr_artem} shows the mass and radius of GJ~357~b compared to various modeled bulk compositions. The planet lies close to the Earth-like density curve, and the $1\sigma$ radius uncertainty does not permit a $>0.01\,\mathrm{wt}\%$ (or equivalently $\gtrsim 180\,\mathrm{bar}$) $\mathrm{H}_2$–He envelope. We can estimate the lifetime of such a $\sim 180\,\mathrm{bar}$ primary atmosphere against escape using an energy-limited XUV (1–90 nm) approximation (e.g., \citealt{Catling_Kasting_2017}):

\begin{equation} \label{eq:escape}
\dot{M}_{\mathrm{atm}} = \eta \frac{\pi R_{\mathrm{p}}^3 L_{\mathrm{XUV}}}{4 \pi a^2 G M_{\mathrm{p}}}.
\end{equation}

Here, $\dot{M}_{\text{atm}}$ is the escape rate in $\mathrm{kg\,s^{-1}}$, $R_{\mathrm{p}}$ is the planet radius in meters, $G$ is the gravitational constant, $M_{\mathrm{p}}$ is the planet mass in kilograms, $L_{\mathrm{XUV}}$ is the stellar XUV luminosity in watts, $a$ is the semi-major axis in meters, and $\eta$ is the escape efficiency, which we assume to be 0.3 \citep{lugerbarnes2015}. We estimate the modern-day XUV luminosity using an XMM-Newton observation \citep{modirrousta-galian_gj_2020} of GJ 357, which recorded an X-ray luminosity of $L_\mathrm{X} = 10^{25.73}\,\mathrm{erg\,s^{-1}}$. Following the relation from \citet{Sanz-Forcada}, we derive an incident XUV flux of $\mathrm{F_{XUV}} = 1.156 \times 10^{2}\,\mathrm{erg\,s^{-1}\, cm^{-2}}$.

Using this incident XUV in Equation \eqref{eq:escape} yields a mass-loss rate of $6.63 \times 10^5\,\mathrm{kg\,s^{-1}}$. Dividing the upper bound $\mathrm{H}_2$–He mass fraction of $0.01\,\mathrm{wt}\%$ by this escape rate suggests that such a low-MMW atmosphere would persist for only about $50\,\mathrm{Myr}$, a short period compared to the planet's age, which has a lower limit of $5\,\mathrm{Gyr}$ \citep{modirrousta-galian_gj_2020}.

In the calculation above, we used an estimate of GJ~357~b's current incident XUV. A primary atmosphere is even less plausible when considering the system's XUV history. During the system's saturation period, which may have lasted several hundred million years \citep{Penz,Sanz-Forcada}, the XUV luminosity would have been even higher, around  $1.131 \times 10^{5}\,\mathrm{erg\,s^{-1}\, cm^{-2}}$ (following \citealt{owen_evaporation_2017}), or three orders of magnitude larger than the present-day value estimated above. Any primary atmosphere that may have accreted onto GJ~357~b during formation would have readily escaped in as little as $500\,\mathrm{kyr}$, assuming an initial H-He rich envelope of around 1.97 wt\% (mass fraction, estimated from \cite{Rogers2023} Equation 2, following \cite{Ginzburg2016}).




These calculations highlight how unlikely it is for GJ 357b to retain a persistent primordial atmosphere. The short atmospheric lifetime, compared to the system's multi-gigayear age, are consistent with our non-detection of a primary atmosphere.

\subsection{Does GJ357b have a high-MMW secondary atmosphere or no atmosphere at all?}

Our observations and the short lifetime of hydrogen against XUV-driven escape (Section \ref{sec:no_primary_atmopshere}) leave two main possibilities for GJ 357b's atmosphere: (1) the planet could possess a relatively thick atmosphere with a high mean molecular weight (MMW), or (2) it may have no atmosphere at all. Figure \ref{fig:atm_examples} shows various high-MMW atmospheres (e.g., $\mathrm{CO}_2$, $\mathrm{CH}_4$, $\mathrm{H}_2\mathrm{O}$, or $1000\times$ solar metallicity) that are allowed by the data. However, given the planet's XUV history, not all of these compositions are equally probable. As we showed in Section \ref{sec:no_primary_atmopshere}, hydrogen and helium would have been rapidly eroded from GJ~357~b. Other volatiles like carbon and oxygen would also have been dragged into space in the escaping hydrodynamic wind, but at much lower rates than H and He \citep{lugerbarnes2015}. The result should be an atmosphere with very few hydrogen-bearing species like $\mathrm{H}_2\mathrm{O}$ and $\mathrm{CH}_4$. Indeed, using a self-consistent model of planetary evolution that includes XUV-driven escape of H, C and O, \citet{Krissansen-Totton} showed that when a primary atmosphere is eroded from a warm rocky world orbiting an M star, the most common outcome is a tenuous O$_2$- or CO$_2$-rich composition. Other similar studies of planetary evolution have come to similar conclusions \citep[e.g.,][]{lugerbarnes2015}. Therefore, while $\mathrm{H}_2\mathrm{O}$, $\mathrm{CH}_4$, and $1000\times$ solar metallicity atmospheres fit our data, they are perhaps less probable than more oxidizing atmospheres (e.g., $\mathrm{O}_2$- or $\mathrm{CO}_2$-rich) when considering planetary evolution.




There is currently theoretical disagreement on the lifetime of high MMW atmospheres against thermal escape, so it is not clear if a O$_2$- or CO$_2$-rich atmosphere would persist on GJ~357~b over geologic time.
The sophisticated upper atmospheric model used in \citet{Looveren2024} suggests that high-MMW atmospheres could experience rapid thermal escape that would likely overwhelm any possible replenishment by volcanism. However, another model presented in \citet{Nakayama2022} disagrees with these results, arguing that atomic line cooling, a process perhaps not fully accounted for in \citet{Looveren2024}, would prevent substantial thermal escape from secondary atmospheres even for XUV irradiation up to 1000$\times$ the modern Earth's (4.64 erg cm$^{-2}$ s$^{-1}$, \citet{lugerbarnes2015}). More recently, \citet{Chatterjee2024} developed an analytical framework to estimate the escape of secondary atmospheres. They argue that \citet{Nakayama2022} did not account for how ion-electron interactions reduce the MMW of the upper atmosphere and, therefore, \citet{Nakayama2022} overestimated the threshold temperature for hydrodynamic escape. Consequently, the new model of \citet{Chatterjee2024} is more pessimistic than \citet{Nakayama2022} finding that an incident XUV of several hundred times Earth's (or less in some cases) can cause the rapid loss of a secondary atmosphere. Note however, that \citet{Chatterjee2024} took an approximate analytical approach for modeling the thermosphere, whereas the \citet{Nakayama2022} model is full-physics and more self-consistent than \citet{Chatterjee2024}. Overall, the thermal atmospheric escape of high MMW atmospheres is an active area of research with evolving conclusions, so it is hard to confidently predict whether GJ~357~b has an atmosphere or if it is instead airless.

\subsection{Future Prospects for Characterizing a Secondary Atmosphere on GJ~357~b}

To further investigate the presence of a secondary atmosphere on GJ~357~b, one option is to collect additional JWST transmission observations with NIRSpec/G395H to increase the data precision and search for a 4.3 $\mu$m CO$_2$ feature. \citet{Taylor2025} recently used \texttt{PandExo} to show that three total G395H transits could detect CO$_2$ to $3\sigma$ confidence in a 1 bar N$_2$ atmosphere with 1000 ppm CO$_2$. The G395H spectrum presented here has errors $\sim50\%$ larger than the values predicted by \texttt{PandExo}, so we expect $>3$ transits would instead be required. Further low-precision transmission observations at wavelengths outside the G395H band pass will not help constrain a secondary atmosphere on GJ~357~b's. Indeed, in addition to performing G395H error calculations, \citet{Taylor2025} presented a NIRISS/SOSS transmission spectrum of the planet. Their observations were featureless, ruling out chemical equilibrium atmospheres with metallicities $<100\times$ solar ($\sim4$ g/mol). Our G395H spectrum improves on these constraints, ruling out metallicities $\leq 300-500 \times$ solar ($\le$ 8 g/mol). A joint analysis of the two datasets may exclude a higher metallicity, although it will not distinguish between the secondary atmosphere and bare rock scenarios because both spectra are flat. 

In summary, future transmission investigations of GJ~357~b should ideally work to increase data precision near the 4.3 $\mu$m CO$_2$ feature rather than expand the wavelength coverage. However, a high precision spectrum may be hard to achieve due to stellar contamination. Even if accurate data is acquired, the tranmission spectrum could be flat and degenerate between a bare rock and a high MMW atmosphere with a high altitude cloud top.






To avoid stellar contamination and cloud degeneracies, the nature of GJ~357~b's atmosphere might instead be investigated by measuring the planet's day side thermal emission with JWST MIRI secondary eclipse observations. Indeed, at the time of writing, there is a single archived but not-yet-published JWST MIRI F1500W secondary eclipse observation of GJ~357~b \citep{DiamondLowe2023}. Here, we show that these new data should provide additional constraints on GJ~357~b's atmosphere, although several additional F1500W eclipses and/or emission observations at other wavelengths will likely be necessary to confidently detect or rule out an atmosphere. Secondary eclipse observations are diagnostic of an atmosphere largely because they are sensitive to the redistribution of incident starlight energy. In the absence of an atmosphere, a deep secondary eclipse (corresponding to a relatively high day-side emission temperature) is expected because absorbed starlight is immediately reradiated. In contrast, a thick atmosphere should reduce the day-side brightness temperature by redistributing heat to the night side with winds, and furthermore, an atmosphere can have characteristic spectral features (e.g., 15 $\mu$m $\mathrm{CO}_2$ absorption).

To assess the value of the existing MIRI F1500W observation, we simulated the expected eclipse precision with the JWST ETC \citep{Pontoppidan2016}. Assuming 3 hours of out-of-eclipse observations and a 1.5 hour eclipse duration, the ETC predicts the planet-to-star flux can be measured with 22 ppm error for a single eclipse. We further inflate this error to 28 ppm, because the F1500W observations of TRAPPIST-1~b had a precision 29\% larger than predicted by the ETC \citep{Greene2023}. 

\begin{figure}
    \centering
    \includegraphics[width=\linewidth]{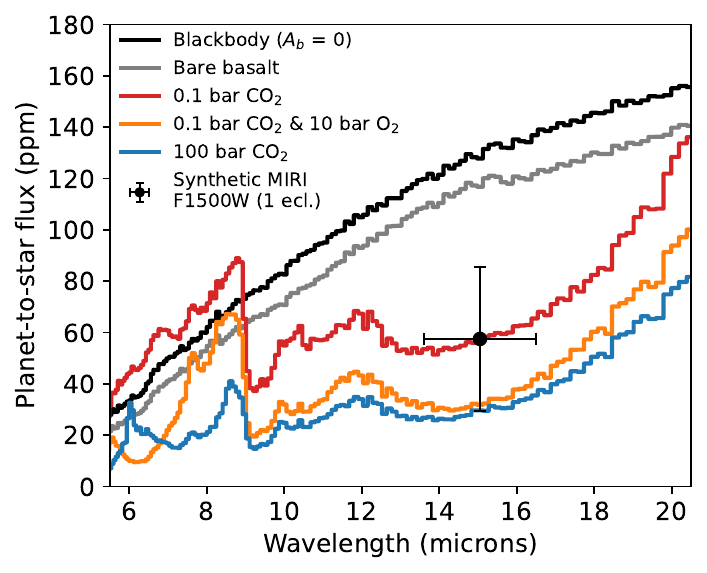}
    \caption{Predicted day side thermal emission from GJ~357~b for a bare rock with zero albedo (black line), bare basalt (gray line), pure CO$_2$ atmospheres with surface pressures of 0.1 and 100 bar (red and blue lines), and an atmosphere with 0.1 bar CO$_2$ and 10 bar O$_2$ (orange line). The black data point is a synthetic JWST MIRI F1500W measurement for a single secondary eclipse. One F1500W eclipse should be able to distinguish bare basalt from a $>0.1$ bar CO$_2$ atmosphere to $\gtrsim 2 \sigma$ confidence, constituting tentative evidence for or against an atmosphere.}
    \label{fig:thermalemission}
\end{figure}

To simulate the day side climate and emission, we use the 1-D climate model in the \texttt{Photochem} software package \citep{Wogan2024photochem}, previously used to simulate the day side emission of L~98-59~c and L~168-9~b \citep{Scarsdale2024,Alam2024}. The code accounts for day-to-night heat redistribution using the \citet{Koll2022} parameterization. Figure \ref{fig:thermalemission} shows the predicted thermal emission of a zero-albedo bare rock, bare basalt (using the \citet{Hu2012} basalt spectrum), pure CO$_2$ atmospheres with surface pressures of 0.1 and 100 bars, and an atmosphere with 0.1 bar CO$_2$ and 10 bar O$_2$. If GJ~357~b has a 0.1 bar CO$_2$ atmosphere, then the currently archived F1500W eclipse should rule out a bare basalt planet with $\sim 2 \sigma$ confidence. For a Venus-like 100 bar CO$_2$ atmosphere or a 10 bar O$_2$ atmosphere with 0.1 bar CO$_2$, the confidence increases to $\sim 3 \sigma$. These atmosphere detection confidences are optimistic because bare rock planets may have surfaces with a higher albedo than basalt \citep{Hammond2025}.

In summary, we find that the existing F1500W eclipse \citep{DiamondLowe2023} may provide tentative evidence for or against a CO$_2$-O$_2$ atmosphere. A more conclusive detection will require further F1500W eclipses to increase the precision and/or perhaps observations at other wavelengths (MIRI LRS or the other MIRI photometry filters) to fully resolve a gas absorption feature, like the 15 $\mu$m CO$_2$ absorption band.

\section{Summary \& Conclusions}
\label{sec:summary}

We have presented JWST NIRSpec G395H transmission spectroscopy observations of the 1.2 R$_{\oplus}$ exoplanet GJ~357~b. Based on our analysis of the transmission spectrum and atmospheric modeling, our key findings are summarized below:

\begin{itemize}
    \item The transmission spectrum of GJ~357~b shows no discernible atmospheric spectral features (Figure \ref{fig:ts}). By comparing the observed spectrum with physical atmospheric models, we rule out atmospheres with mean molecular weights (MMWs) $\le$ 8 g/mol and metallicities $\leq 300-500 \times$ solar. This suggests the absence of a low-MMW primary atmosphere.
    \item The lack of a primary atmosphere is consistent with GJ~357~b's high bulk density and the short atmospheric lifetime of hydrogen to escape. Energy-limited escape calculations indicate that the current, most massive primary H$_2$-He envelope allowed by the planet's density would be lost within 50 million years, much shorter than the planet's estimated age of over 5 billion years.
    \item Considering the planet's XUV history and evolutionary models in the literature, two scenarios remain: GJ~357~b may possess a high-MMW secondary atmosphere, perhaps rich in oxidizing gases $\mathrm{CO}_2$ or $\mathrm{O}_2$, or it may have no atmosphere at all. Hydrogen-bearing species like $\mathrm{H}_2\mathrm{O}$ and $\mathrm{CH}_4$ are less probable due to the preferential loss of hydrogen during escape.
    \item Even high-MMW atmospheres may not be stable over geologic timescales due to ongoing atmospheric escape processes, including non-thermal escape mechanisms like ion pickup and sputtering. While volcanic outgassing could replenish the atmosphere, models disagree on the efficiency of atmospheric loss for high-MMW atmospheres.
    \item The archived but yet-to-be-published JWST MIRI F1500W secondary eclipse observation of GJ~357~b \citep{DiamondLowe2023} should provide tentative evidence for or against a CO$_2$-O$_2$ atmosphere with $\gtrsim 0.1$ bar CO$_2$. Conclusively detecting or ruling out an atmosphere will require further F1500W eclipses and/or observations at additional wavelengths.

\end{itemize}

Our findings contribute to the growing body of evidence that warm, rocky exoplanets orbiting M dwarfs may lack substantial atmospheres, highlighting the importance of atmospheric escape processes in shaping planetary atmospheres. Future observations, particularly thermal emission measurements, will be crucial in determining the presence and composition of any remaining atmosphere on GJ~357~b.

\section{Acknowledgements}

\noindent The data products for this manuscript can be found at the following Zenodo repository: \dataset[doi: 10.5281/zenodo.15742733]{https://doi.org/10.5281/zenodo.15742733}. This work is based on observations made with the NASA/ESA/CSA James Webb Space Telescope. The data were obtained from the Mikulski Archive for Space Telescopes at the Space Telescope Science Institute, which is operated by the Association of Universities for Research in Astronomy, Inc., under NASA contract NAS 5-03127 for JWST. These observations are associated with program \#2512. Support for program \#2512 was provided by NASA through a grant from the Space Telescope Science Institute, which is operated by the Association of Universities for Research in Astronomy, Inc., under NASA contract NAS 5-03127. This work is funded in part by the Alfred P. Sloan Foundation under grant G202114194. Support for this work was provided by NASA through grant 80NSSC19K0290 to JT and NLW. This work benefited from the 2022 and 2023 Exoplanet Summer Program in the Other Worlds Laboratory (OWL) at the University of California, Santa Cruz, a program funded by the Heising-Simons Foundation. N.F.W was supported by the NASA Postdoctoral Program.

Co-Author contributions are as follows: JAR led the data analysis and write up of this work. NFW led the modeling efforts. NLW provided a secondary independent reduction and analysis of the data. NFW and AA aided in the theoretical interpretation of the data. MA and JK advised JA throughout the analysis of this work.

We thank the referee for their thorough comments that improved the paper.

\software{\texttt{astropy} \citep{Astropy2013, Astropy2018, AstropyCollaboration2022}, \texttt{batman} \citep{Kreidberg2015}, \texttt{emcee} \citep{Foreman-Mackey2013}, \texttt{Eureka!} \citep{2022JOSS....7.4503B}, \texttt{ExoTiC-JEDI} \citep{Alderson2022JEDI}, \texttt{ExoTiC-LD} \citep{Grant2022}, \texttt{Matplotlib} \citep{matplotlib},  \texttt{NumPy} \citep{numpy}, \texttt{PandExo} \citep{Batalha2017}, \texttt{PICASO} \citep{Batalha2018, Mukherjee2023}, \texttt{photochem} \citep{Wogan2023}, \texttt{scipy} \citep{scipy}, STScI JWST Calibration Pipeline \citep{Bushouse2022}, \texttt{ultranest} \citep{Ultranest}, \texttt{xarray}} 

The JWST data presented in this article were obtained from the Mikulski Archive for Space Telescopes (MAST) at the Space Telescope Science Institute. The specific observations analyzed can be accessed via \dataset[doi: 10.17909/9yy2-vw75]{https://doi.org/10.17909/9yy2-vw75}.

\bibliography{references}{}
\bibliographystyle{aasjournal}

\end{document}